\documentclass{iopart}
\usepackage{dcolumn}
\usepackage{bm}
\usepackage{graphicx}
\usepackage{hyperref}
\usepackage{mathptmx}
\usepackage{subfigure}
\usepackage[utf8]{inputenc}
\usepackage{bbold}
\usepackage{color}

\makeatletter

\renewcommand{\vec}[1]{\mathbf{#1}}

\newcommand{\eqref}[1]{(\ref{#1})}

\begin{document}

\title{Efficient algorithms for general periodic Lorentz gases in two and three dimensions}

\author{Atahualpa S.~Kraemer$^{1}$, Nikolay Kryukov$^{2}$ and David P.~Sanders$^{2}$}

\address{$^{1}$ Institut für Theoretische Physik II - Soft Matter
Heinrich-Heine-Universität Düsseldorf
Building 25.32
Room O2.56
Universitätsstrasse 1
D-40225 Düsseldorf, Germany\\
$^{2}$Departamento de F\'isica, Facultad de Ciencias, Universidad Nacional
Aut\'onoma de M\'exico,
Ciudad Universitaria, M\'exico D.F.\ 04510, Mexico
}

\eads{\mailto{ata.kraemer@gmail.com}, \mailto{kryukov@ciencias.unam.mx}
and 
  \mailto{dpsanders@ciencias.unam.mx}} \date{\today}

\begin{abstract}

We present efficient algorithms to calculate trajectories for periodic Lorentz gases consisting of square lattices of 
circular obstacles in two dimensions, and simple cubic lattices of spheres in three dimensions; these become increasingly efficient as the radius of the ostacles tends to $0$, the so-called Boltzmann--Grad limit.
The 2D algorithm applies continued fractions to obtain the exact disc with which a particle will collide at each step, instead of using periodic boundary conditions as in the classical algorithm. The 3D version incorporates the 2D algorithm by projecting to the three coordinate planes. 
As an application, we calculate distributions of free path lengths close to the Boltzmann--Grad limit for certain Lorentz gases.
We also show how the algorithms may be applied to deal with general crystal lattices. \end{abstract}

\maketitle

\section{Introduction}

Lorentz gases are simple physical systems that present deterministic chaos \cite{cvitanovic1992investigation}, and are a popular model in statistical mechanics and nonlinear dynamics. This model consists of point particles that move freely until they encounter obstacles, often spheres, where they undergo elastic collisions. 
%When these obstacles are arranged periodically at the vertices of a grid (``periodic Lorentz gas''), the model is equivalent to a Sinai billiard \cite{bunimovich1981statistical}.

These systems can have different configurations of obstacles, e.g., random arrangements \cite{latz1997lyapunov,dellago1997lyapunov, van1998chaotic} or quasiperiodic structure \cite{kraemer2013embedding,wennberg2012free}. However, due to its simplicity, the periodic case has been most widely studied; see, e.g., \cite{bunimovich1981statistical,bleher1992statistical, chernov1994statistical, gilbert2011diffusive}. In this case, the model is equivalent to a Sinai billiard  \cite{bunimovich1981statistical}. Many of the results obtained theoretically for these gases are in the limit where obstacles are very small, i.e., the so-called Boltzmann--Grad limit \cite{caglioti2003distribution, golse2012recent,boca2007distribution,golse2006periodic,caglioti2008boltzmann,caglioti2010boltzmann, golse2000distribution,marklof2008kinetic, bourgain1998distribution}. There are still many interesting open questions in this area
\cite{gilbert2009persistence,marklof2011periodic,nandori2014tail, dettmann2012new}.

The standard simulation method for periodic Lorentz gases is to reduce to a single cell with periodic boundary conditions, and, in the simplest case, an obstacle in the centre of the cell
\cite{sanders2005fine, sanders2008normal}. However, this requires that the program check in each cell whether the particle collides with the obstacle in the cell, or if it will move to the next cell. 
%That implies solving at least two linear equations and one quadratic, then checking if the solution of the quadratic equation is real and finally taking the maximum value of the two or three results. 
If the obstacle is large, it is quite likely that the particle will collide each time it crosses into a new cell. However, for very small obstacles, this method becomes very inefficient.

Instead, we would like to just find the coordinates of the next obstacle with which the particle will collide, given its initial position and velocity.
%As we will see in section \ref{algorithm} this is indeed the first solution of the Diophantine inequality:
%
%\begin{equation}
%|\alpha q+b-p|\leq  \frac{v^2}{v_x} r
%\end{equation}
%
%where $r$ is the radius of the obstacles, $q$ and $p$ are integer variables, and $\alpha$ and $b$ are two real numbers related to the position $\vec{x}$ and velocity $\vec{v}$ of the particle. 
This turns out to be  closely related to the best rational approximant to an irrational number, and  can be solved using the continued-fraction algorithm. 
 Continued fractions have often been used to provide information about the free path distribution of the periodic Lorentz gas in the Boltzmann--Grad limit \cite{caglioti2003distribution, golse2012recent,boca2007distribution,golse2006periodic,caglioti2008boltzmann, caglioti2010boltzmann, golse2000distribution, bleher1992statistical,chernov1994statistical}.
  An algorithm along these lines was previously developed: see comments in \cite{zacherl1986power}; however, it was never published %\footnote{T.~Geisel, private communication}.
[T.~Geisel, private communication].
 Caglioti and Golse developed a method to encode the trajectories of particles using the continued fraction algorithm and the so-called 3-length theorem \cite{caglioti2003distribution, golse2012recent}.

However, Golse's algorithm works only if the particle leaves the surface of a disk. This restriction prevents the algorithm from being used in other geometries, such as two incommensurate overlapping arrays of square lattices, or with different shapes of obstacles; such systems may produce a number of surprising effects \cite{marklof2014power}.  

On the other hand, due to the construction of Golse's algorithm, it is not possible to use it in higher dimensions, which is ``a notoriously more difficult problem'' \cite{golse2012recent}.
Recent advances on multidimensional continued-fraction algorithms  may provide a possible future direction \cite{khanin_multidim_cont_frac, lagarias_multidim_cont_frac}, although here we have opted for a different approach for higher-dimensional systems.

In this paper, we develop an efficient algorithm to find a collision with a 2D square lattice of discs starting from an arbitrary initial condition. We then use that 2D algorithm as part of an efficient algorithm for a 3D simple cubic lattice by projecting onto coordinate planes. Finally, we show how obstacles arranged on arbitrary (periodic) crystal lattices may be treated.

\section{Classical algorithm for the periodic Lorentz gas}

\begin{figure}
\centering
\includegraphics*[width=200pt]{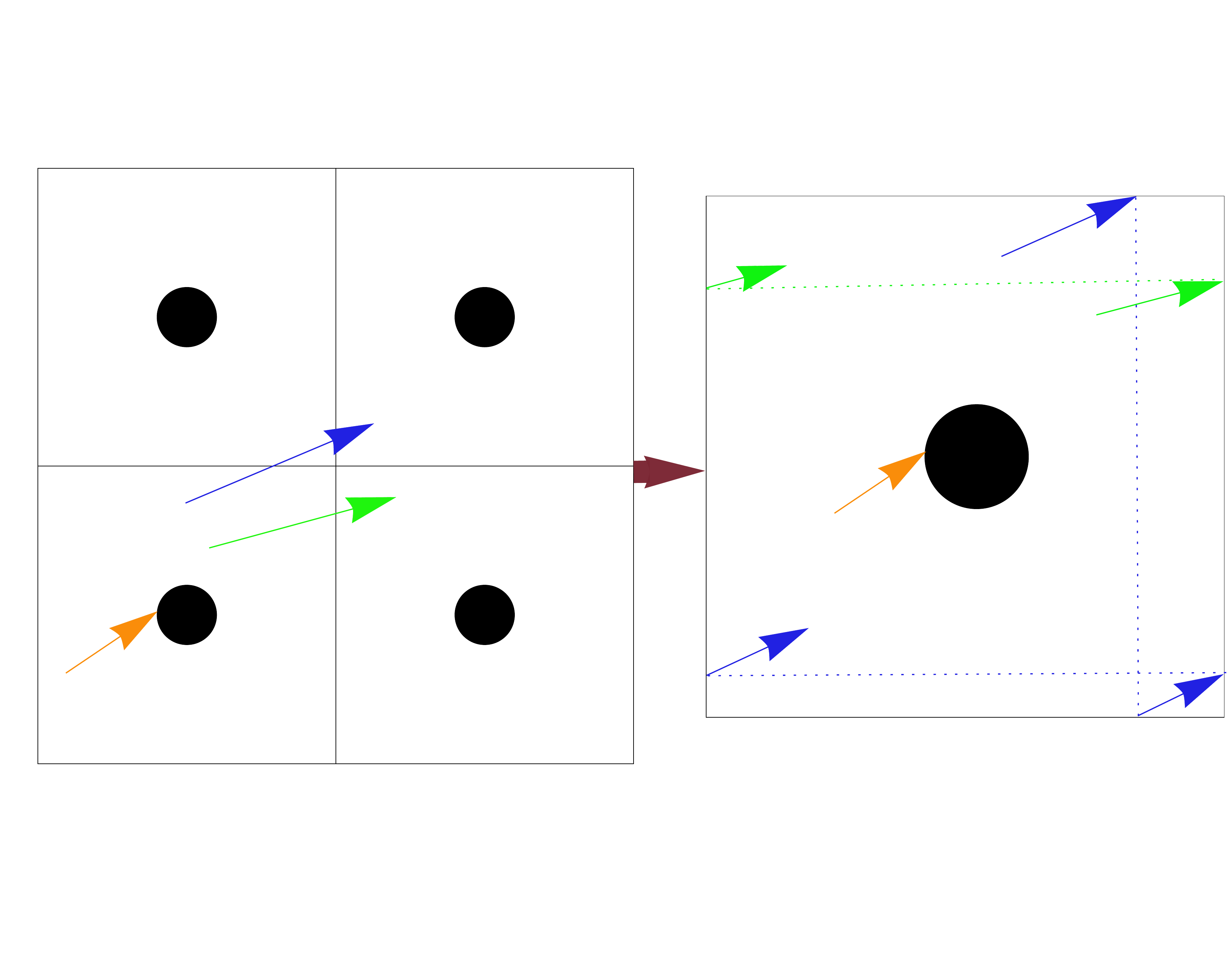}
\caption{Reducing the dynamics in a periodic lattice to a single cell with periodic boundary conditions.}
\label{fig:classical}
\end{figure}

We begin by recalling the classical algorithm for a Lorentz gas on a $d$-dimensional (hyper)-cubic lattice, where each cell contains a single spherical obstacle of radius $r$. The simplest method is to locate the centre of the obstacle at the centre of a cubic cell $[-\frac{1}{2}, \frac{1}{2})^d$, and to track which cell $\vec{n} \in \mathbb{Z}^{d}$ the particle is in using periodic boundary conditions: when a particle hits a cell boundary, its position is reset to the opposite boundary and the cell counter $\vec{n}$ is updated accordingly; see Figure~\ref{fig:classical}.

%\begin{equation}
%\vec{x} = \vec{x_0} - \vec{n} ,
%\end{equation}
%where $\vec{x_0}$ is the original position of the particle and $\vec{n}$ are the integer coordinates of the centre of the unit cell with respect to the origin:
%\begin{equation}
%n_i = \lfloor x_{0,i} - \textstyle \frac{1}{2} \rfloor.
%\end{equation}

In each cell, the classical algorithm is as follows. 
For a particle with initial position $\vec{x}$ and velocity $\vec{v}$, a collision occurs with the disc with centre at $\vec{c}$ and radius $r$ at a time $t^{\ast}$ if
\begin{equation}
\| \vec{x} + \vec{v} t^{\ast} - \vec{c} \| = r.
\end{equation}
This gives a quadratic equation for the collision time, and hence
\begin{equation}
t^{\ast} = -B - \sqrt{B^2 - C}
\label{eq:collision_time}
\end{equation}
where
\begin{equation}
B= \frac{(\vec{x}  -\vec{c}) \cdot \vec{v}}{v^2}; \qquad
C= \frac{(\vec{x} - \vec{c})^2 -r^2}{v^2},
\end{equation}
provided that the condition $B^2 - C \ge 0$ is satisfied. If this happens, then the collision position is $\vec{x} + \vec{v} t^{\ast}$. If the condition is not satisfied, then the trajectory misses the disc.

If no collision with the obstacle occurs, i.e.\ when $B^{2} - C < 0$, the velocity is conserved and the particle will hit one of the cell boundaries. To determine which boundary will be hit, we find intersection times of the trajectory with each cell boundary (lines 2D or planes in 3D), given by
$$t_{i, \pm} = \frac{\pm \frac{1}{2} - x_i}{v_i},$$
where $i$ runs from 1 to the number of dimensions (2 or 3) and the sign corresponds to the two opposite faces in direction $i$. The least positive time then gives the collision time with the boundary. Depending on which boundary was hit, we move to the new unit cell and repeat the process: if $t_{i,\pm}$ is the minimum time, then the positive (resp.~negative) $i$th boundary is hit, and the $i$th component of the cell is updated to $n_{i}' = n_{i} \pm 1$.

\section{Efficient 2D algorithm}

The classical algorithm is efficient for large radii $r$, but very inefficient once $r$ is small, since a trajectory will cross many cells before encountering a disc.

In this section, we develop an algorithm to simulate the periodic Lorentz gas on a 2D square lattice, based on the use of continued fractions, whose
 goal is to calculate \emph{efficiently} the first disc hit by a particle, even for very small values of the radius $r$. Without loss of generality, we will use the lattice formed by the integer coordinates in the 2D plane. 
 
\begin{figure}
\centering
\includegraphics*[width=200pt]{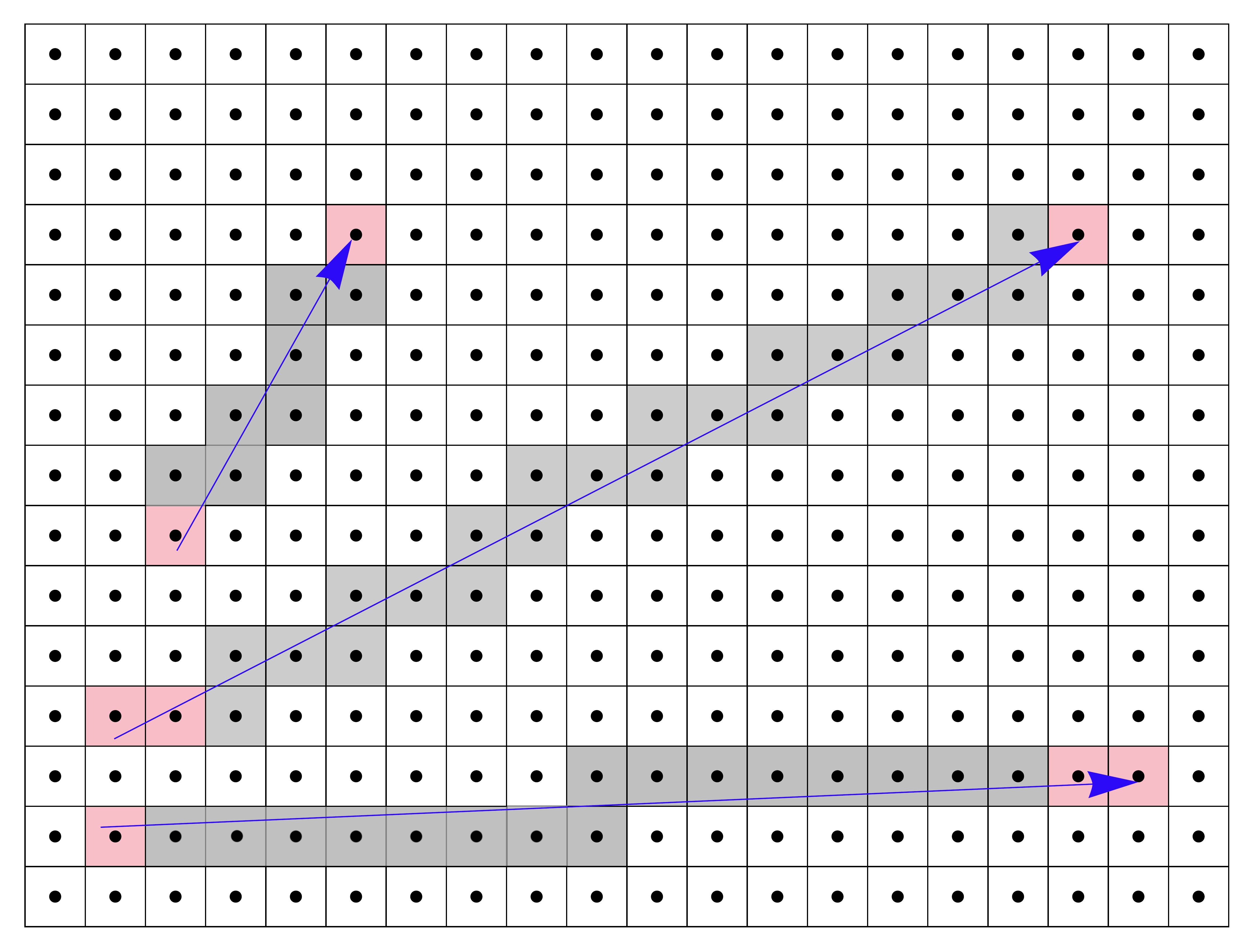}
\caption{Cells covered by the classical algorithm, compared to the few steps required by the efficient algorithm.}
\label{fig:efficient}
\end{figure}

We wish to calculate the minimal time $t^{\ast}>0$ such that a collision occurs with some disc centred at $\vec{c} = (q,p)$, with $q$ and $p$ integers, by ``jumping'' straight to the correct disc; see Figure~\ref{fig:efficient}.
%\begin{equation}
%\|\vec{x_0}+\vec{v_0} \, t^{\ast} -  \vec{c} \| = r,
%\label{eq:classical}
%\end{equation}
%where $r$ is the radius of the obstacle, and $\vec{x_0}$ and $\vec{v_0}$ are the initial position and velocity of the particle. The collision point is then
%$\vec{x_0}+\vec{v_0} \, t^{\ast}$.

\subsection{Continued fraction algorithm: approximation of an irrational number by a rational}

In this section we recall the continued fraction algorithm and some properties of continued fractions; see, e.g., 
\cite{niven2008introduction} for proofs. The geometrical interpretation has been suggested before by many other authors; see, for example, \cite{nogueira1995three}. 

A continued fraction is obtained via an iterative process, representing a number $\alpha$ as the sum of its integer part, $a_0$, and the reciprocal of another number, $\alpha_1:=\alpha-a_0$, then writing $\alpha_1$ as the sum of its integer part, $a_1$, and the reciprocal of $\alpha_2:=\alpha_1-a_1$, and so on. This gives the continued fraction representation of $\alpha$:
\begin{equation*}
  \alpha = a_0 + \frac{1}{\displaystyle a_1
          + \frac{1}{\displaystyle a_2
          + \frac{1}{\displaystyle a_3 + \dots}}}
\end{equation*}

This iteration produces a sequence of integers $\lfloor \alpha \rfloor=a_0$, $\lfloor \alpha_1 \rfloor=a_1$, $\lfloor \alpha_2 \rfloor=a_2$, etc. 
We define inductively two sequences of integers $\{ p_n\}$ and $\{ q_n\}$ as follows:
\begin{eqnarray}
p_{-2} = 0;  \quad p_{-1} = 1;  \quad p_i &=a_i p_{i-1}+p_{i-2};
\label{eq:sucesion1}
\\ 
q_{-2} = 1,  \quad q_{-1} = 0,  \quad q_i &=a_i q_{i-1}+q_{i-2}.
\label{eq:sucesion2}
\end{eqnarray}

With this sequence we can approximate any irrational number $\alpha$ using the Hurwitz theorem:
For any irrational number, $\alpha$, all the relative prime integers $p_n$, $q_n$ of the sequences defined in equations \eqref{eq:sucesion1} and \eqref{eq:sucesion2} satisfy 
\begin{equation}
|\alpha- \frac{p_n}{q_n}|\leq  \frac{1}{{q_n}^2}.
\end{equation}

\subsection{Collision with a disc}
The classical algorithm finds the intersection between a line, corresponding to the trajectory of 
the particle, and a circle, corresponding to the circumference of the disc, by solving the quadratic equation~\eqref{eq:collision_time} for $t^{\ast}$. A first improvement follows from observing that we may instead look for the intersection of the trajectory with another line, as follows.  In the following, we take $v_1>0$ and $v_2>0$; by symmetry of the system, we can always rotate or reflect it such that these conditions are satisfied.

%We write the equation of the particle's trajectory as $y=\alpha x+b$, with $\alpha = v_{2} / v_{1}$, and look for its intersection with the vertical line $x=q$ passing through the disc at $(q,p)$. As shown in figure~\ref{fig:circle}, if $|\alpha q+b-p| = |b'| < \delta := r/v_1$, then a collision with the disc $(q,p)$ will occur. Due to the periodic boundary conditions, we can take $0 < b' < 1$, so we need only solve $b<\delta$. 
%This gives us a tool to calculate with which obstacle the particle will collide. 

\begin{figure}
\centering
\includegraphics [width=240pt]{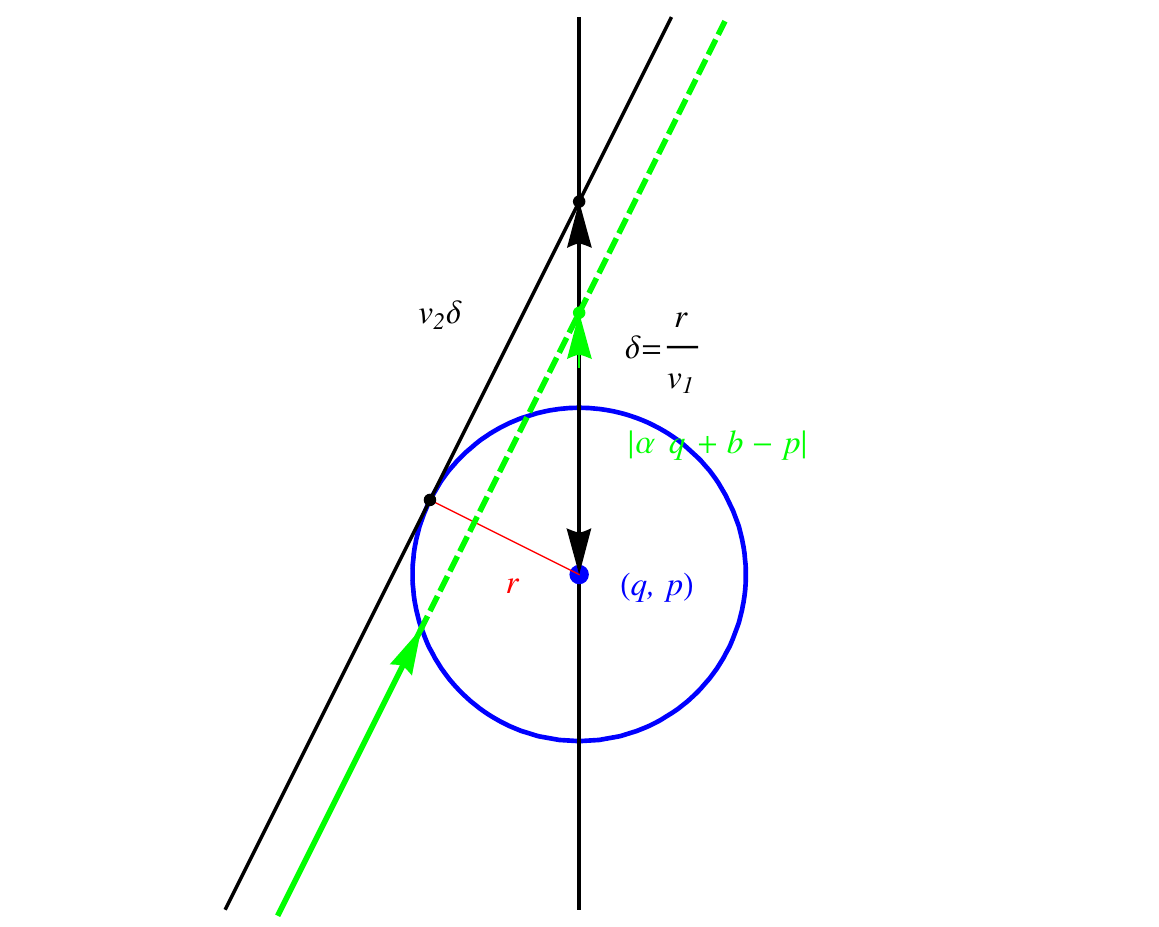}
\caption{Relation between the intersection of a line and a circle with integer coordinates and the intersection of the line $x = q$. }
\label{fig:circle}
\end{figure}

We write the equation of the particle's trajectory as $y=\alpha x+b$, with slope $\alpha = v_{2} / v_{1}$, and look for its intersection with the vertical line $x=q$ passing through the disc at $(q,p)$. As shown in figure~\ref{fig:circle}, if $|\alpha q+b-p| < \delta := r/v_1$, then a collision with the disc $(q,p)$ will occur. Due to the periodic boundary conditions, we can redefine $b := \{|\alpha q+b-p|\}$ where $\{\cdot\}$ denotes the fractional part. Thus, $0 < b < 1$, and we need only solve $b<\delta$.

We do not need to apply periodic boundary conditions at every step; rather, we only need to check 
\begin{equation}
|\{ \alpha  q_n \}+b -1|< \delta, 
\label{eq:master}
\end{equation}
where $\{ \cdot \}$ denotes the fractional part, and $q_n=q_{n-1}+1$, where $q_1$ is the $x$-coordinate of the closest obstacle to the particle at $t=0$. Then, the first $q_n$ that satisfies this inequality will be $q$. To calculate $p$, we use that either $p=\lfloor \alpha q +b\rfloor$ or $p=\lfloor \alpha q +b\rfloor+1$.

Now, to simplify the algorithm further, consider the integer coordinates $(q_n, p_n)$ such that
\begin{equation}
|\alpha q_n -p_n + b|< \delta,
\label{eq:1}
\end{equation}
and for any pair of numbers $(i,j)$ such that $i<q_n$, then $|\alpha i -j+ b|> \delta$,  $q=q_n$, and $p=p_n$. 
 
But $|\alpha q_i - p_i + b|$ are the distances between the integer coordinates $(q_i, p_i)$ and the point $( q_i ,\alpha q_i + b)$. Thus, we would like a sequence such that  
\begin{equation}
|\alpha q_i - p_i + b|<|\alpha q_{i-1} - p_{i-1} + b|
\label{eq:iteration}
\end{equation}
for every integer $i>1$. Also, the first pair of integer coordinates $q_0$ and $p_0$ should be $(0, 0)$ or $(0, 1)$, minimizing $| \alpha q_0 - p_0 + b |$, that is

\begin{eqnarray}
%|\alpha q_1 -p_1 + b|< f(b) = \begin{cases} b &\mbox{if } b < 1/2 \\ 
%1-b & \mbox{if } b > 1/2 \end{cases} .
|\alpha q_1 -p_1 + b|< f(b) =&
  \left\{ 
  \begin{array}{@{}l@{\quad}l}
      b,      & \mbox{if } b < \frac{1}{2} \\[\jot]
      1 - b, & \mbox{if } b > \frac{1}{2}.
   \end{array}
   \right.
\label{eq:prima}
\end{eqnarray}

Note that if $b < 1/2$, we have
$p_n= \lfloor \alpha q_n +b  \rfloor= \lfloor \alpha q_n  \rfloor $, if $b+\alpha q_n-\lfloor \alpha q_n  \rfloor < 1$, and  
$p_{n} = \lfloor \alpha q_n  \rfloor+1$, if  $b+\alpha q_n-\lfloor \alpha q_n  \rfloor > 1$.
Whereas if $b>1/2$, we  have $p_n= \lfloor \alpha q_n +b  \rfloor+1= \lfloor \alpha q_n  \rfloor+1$, if $b+\alpha q_n-\lfloor \alpha q_n  \rfloor < 1$ and 
$ p_{n} = \lfloor \alpha q_n  \rfloor+2$, if $b+\alpha q_n-\lfloor \alpha q_n  \rfloor > 1$. Substituting these four cases  in the two cases of equation \eqref{eq:prima}, we obtain that indeed $p_1= \lfloor \alpha q_1  \rfloor+1$. Iterating the inequality \eqref{eq:iteration} we obtain 
\begin{equation}
 p_n= \lfloor \alpha q_n  \rfloor+1.
\label{eq:hn}
\end{equation}
Combining the inequality \eqref{eq:1} with equation \eqref{eq:hn}, we obtain again equation \eqref{eq:master}. 

Thus, we have reduced the solution from two linear equations and one quadratic to one linear equation.
Furthermore, now we do not check in every periodic cell, because if $\alpha >1$,
for every $q_n$ we advance $(\lfloor q_{n}\alpha \rfloor -\lfloor q_{n-1}\alpha \rfloor)$ cells. 
And we do not need to apply periodic boundary conditions until we reach the obstacle. 

\subsection{The Diophantine inequality: $|\alpha p - q|\leq \delta$}

Now, a better algorithm should find a way to find the set of $q_i$, such that inequality \eqref{eq:iteration} holds for every $i$, and there is no integer $q$  such that $q_i<q<q_{i-1}$ for some $i$ and
$|\{ \alpha  q_i \}+b -1|<|\{ \alpha  q \}+b -1| <|\{ \alpha  q_{i-1} \}+b -1|$. 

In order to do this, we can use the continued fraction algorithm to obtain solutions to the inequality $|\alpha q - p|\leq \delta$. This algorithm already gives a sequence of $(q_n,p_n)$ such that $|\alpha q_i - p_i|<|\alpha q_{i-1} - p_{i-1}|$ if $q_{i-1} <q_i$. In addition, the convergents of the continued fractions provide best approximants and hence the smallest solution of the inequality \eqref{eq:prima}.
So, if we turn our inequality \eqref{eq:prima} into this other inequality, we will find our algorithm just by using the continued fraction algorithm.
Indeed, using equation \eqref{eq:hn} and the inequality \eqref{eq:prima}, we obtain $|\{ \alpha  q_1 \} -1|< 2b$ if $b < 1/2$ or $<  2(1-b)$  if $b > 1/2$, 
which is almost the continued fraction inequality, except that $p_1$ is always equal to $\lfloor \alpha q_1  \rfloor+1$. 

%\begin{equation}
%|\alpha  q_1-p_1|< \begin{cases} 2b &\mbox{if } b < 1/2 \\  2(1-b) & \mbox{if } b > 1/2 \end{cases}
%\label{eq:master2}
%\end{equation}

\begin{eqnarray}
    | \alpha q - p | < 
 \left\{ 
  \begin{array}{@{}l@{\quad}l}
 & 2b, \quad \mbox{if } b < \frac{1}{2} \\ 
& 2(1-b), \quad  \mbox{if }  b > \frac{1}{2}.
   \end{array}
   \right.
\label{eq:master2}
\end{eqnarray}

Now we can apply the continued fraction algorithm to obtain $p_1$ and $q_1$ of inequality \eqref{eq:master2}.  If ${\alpha q_1+b} < \delta$ or $1-{\alpha q_1+b} < \delta$, then we have found the center of the obstacle at $(p_1,q_1)$, with  $p_1=\lfloor \alpha q_1  \rfloor+1$; otherwise, we have not found it, but we know that if the center of collision is at $(p,q)$ then
$p\geq p_1$, and $q \geq q_1$. Hence, we can just use $(q_1,p_1)$ even if they do not satisfy inequality \eqref{eq:iteration}. Redefining $b_i$ as $b_0=b$, $b_i=\{\alpha q_i+b\}$, we can calculate a succession of $(p_i,q_i)$. If $b_n<\delta$ the algorithm stops, and the collision will take place with the obstacle centred at the coordinates $(q_n,p_n)$.
Otherwise, if $b_n = b$, then the particle has a rational slope equal to that of a channel, and so is travelling along and parallel to that channel, and hence will never undergo another collision with an obstacle. In this case, the algorithm throws an exception.

\subsection{Complete 2D algorithm}

We now have the necessary tools to implement the algorithm. Pseudo-code for the complete efficient 2D algorithm is given in the Appendix;  source code for our implementation, written in the Julia programming language
\cite{JuliaLang, JuliaArticle},
may be found in the supplementary information.

The functions described above work only for velocities in the first octant, i.e. such that $0 < v_{2} < v_{1}$. If the velocity does not satisfy this condition, we use the symmetry of the system, applying rotations and reflections and then, after obtaining the 
coordinates of the collision, use the inverse transformations to return to the original system; see the Appendix for details.

Finally, to calculate the exact collision point, we use the classical algorithm to obtain the intersection between a line and a circle, and from there the resulting post-collision velocity.

\section{Efficient 3D algorithm}

We now develop an efficient algorithm for calculating the next collision with a sphere in 3D on a simple cubic lattice, which again is designed to be efficient for a small radius $r$. The algorithm works by projecting the geometry onto the 2D coordinate planes and then using the above efficient 2D algorithm in each plane, as follows.

Suppose we project a particle trajectory in a 3D lattice onto one of the $x$--$y$, $x$--$z$ or 
$y$--$z$ planes. We will obtain a periodic square lattice with a 2D trajectory. This trajectory is \emph{not} a trajectory of the 2D Lorentz gas, however -- it may pass through certain discs as if they were not there, and may have non-elastic reflections with other discs. Furthermore, the speed varies.

However, we will use this to apply the 2D algorithm for the projections in each plane, in order to obtain coordinates of a disc in each of the three planes at which the first collision in that plane is predicted to occur.
We now check whether the obstacle coordinates in these projections correspond to the \emph{same} 3D obstacle, i.e. if the coordinates coincide pairwise. If not, then we have not found a true collision in 3D.
We move the particle to the cell containing the obstacle that is furthest away, i.e., has the maximum collision time in its respective plane, and continue.

\begin{figure}
\centering
\includegraphics [width=240pt]{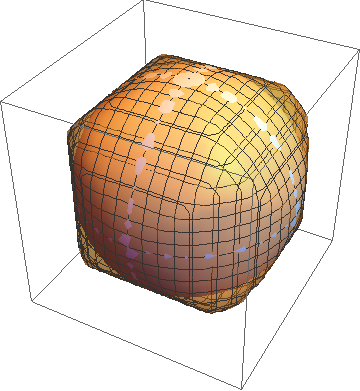}
\caption{A sphere of radius $r$ embedded into the intersection of three orthogonal cylinders of the same radius. The volume inside the intersection but outside the sphere is the region where the 3D algorithm predicts false collisions.}
\label{fig:collision}
\end{figure}

If the obstacle coordinates do coincide pairwise, then this algorithm predicts that there is a collision. However, this may not be true, due to the geometry, as follows.
Calculating a collision with a disc in one of the planes $x$--$y$, $x$--$z$ or $y$--$z$ is equivalent to calculating a collision in space with a cylinder orthogonal to that plane. Joining these coordinates together means calculating a collision with the intersection of three orthogonal cylinders with the same radius. 
Figure~\ref{fig:collision} shows such an intersection of three cylinders, called a tricylinder or Steinmetz solid \cite{tricylinder1974}, together with a sphere of the same radius. The sphere is contained inside the intersection of the cylinders, and has a smaller volume.
 
Thus the algorithm may predict a false collision -- with the tricylinder -- even though the particle does not collide with the sphere. To control this, we check if the particle really does collide with this obstacle by using the classical algorithm; if so, then we have found a true collision, and if not, we move the particle to the next cell and continue applying the algorithm. 
Numerically, we find the probability of a false collision to be around $0.17$. 
% Analytically, the sphere embedded in the tricylinder has the volume $V_{sph} = \frac{4}{3} \pi r^3$, while the tricylinder has the volume $V_{tc} = (16 - 8 \sqrt{2}) r^3$ \cite{tricylinder1974}. The probability of a false collision is therefore
% 
%$$ p_{false} = \frac{V_{tc} - V_{sph}}{V_{tc}} = 1 - \frac{\pi}{6 (2 - \sqrt{2})} \approx 0.106 $$

%This algorithm may be generalised to a hyper-cubic lattice of hyper-spheres in $d$ dimensions.

Pseudo-code for the efficient 3D algorithm is given in the Appendix.

\section{Numerical measurements}

\subsection{Execution time} 
In order to test the efficiency of our algorithms, we measure the average execution time of the function that finds the first collision, starting from an initial point near the origin,
 as a function of obstacle radius, for both the classical and efficient algorithms, in 2D and 3D.

\begin{figure}
\centering
\includegraphics [width=260pt]{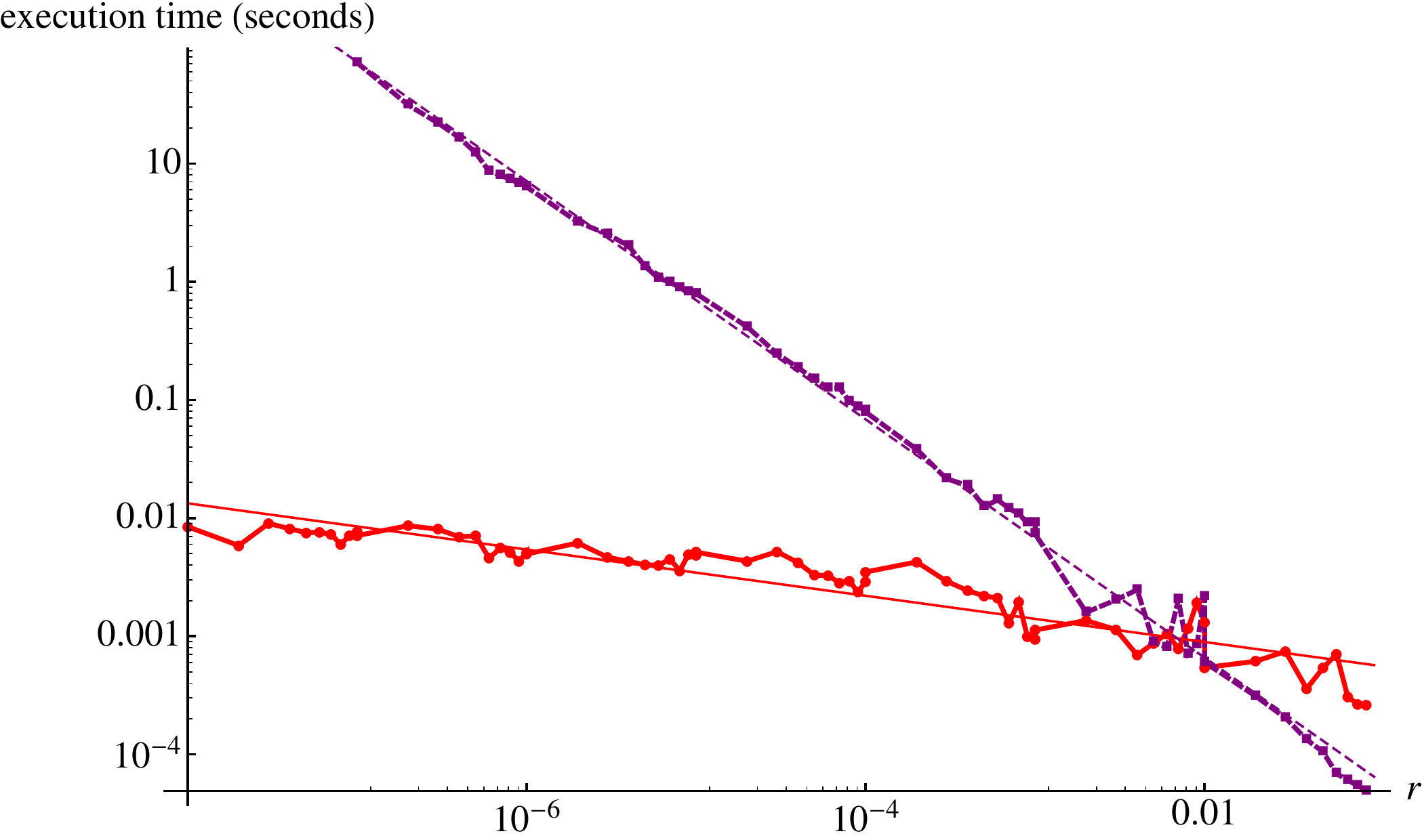}
\caption{Mean execution time to find the first collision in the 2D square Lorentz gas, for the classical (dotted curve) and efficient (solid curve) algorithms. The straight lines show power-law fits. }
\label{fig:time_2D}
\end{figure}

\begin{figure}
\centering
\includegraphics [width=260pt]{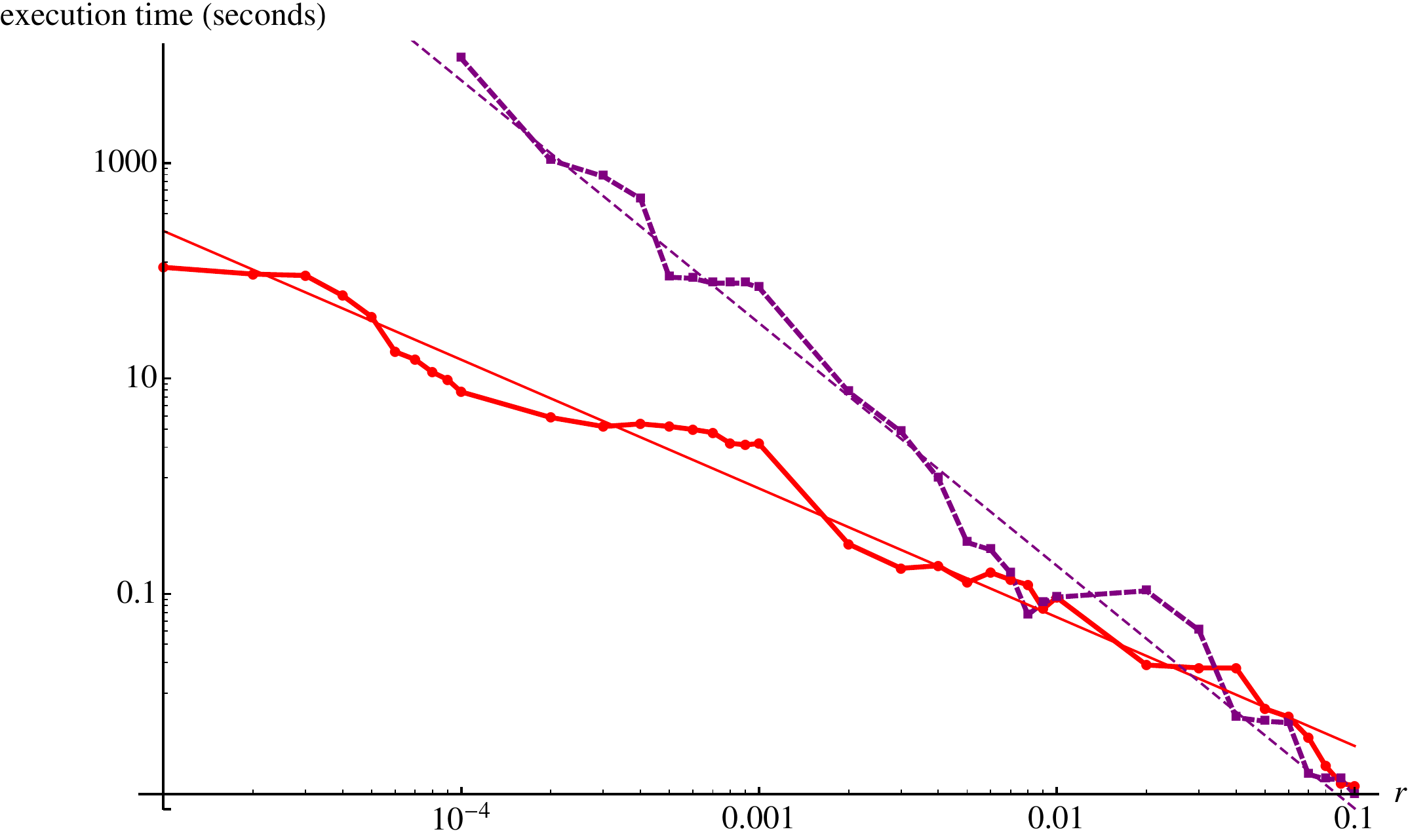}
\caption{Mean execution time to find the first collision in the 3D simple cubic Lorentz gas, for the classical (dotted curve) and efficient (solid curve) algorithms. The straight lines show power-law fits.}
\label{fig:time_3D}
\end{figure}

Figures \ref{fig:time_2D} and \ref{fig:time_3D} show the results for the 2D and 3D algorithms, respectively. We performed power-law fits for the execution time as a function of obstacle radius. For the 2D case, we find an exponent of $-1.01$ for the classical algorithm and $-0.20$ for the efficient algorithm. For the 3D case, the exponents are $-2.25$ and $-1.20$ for classical and efficient, respectively. As we can see, our algorithms are increasingly more efficient for $r < 0.01$. 

\begin{figure}
\centering
\includegraphics [width=260pt]{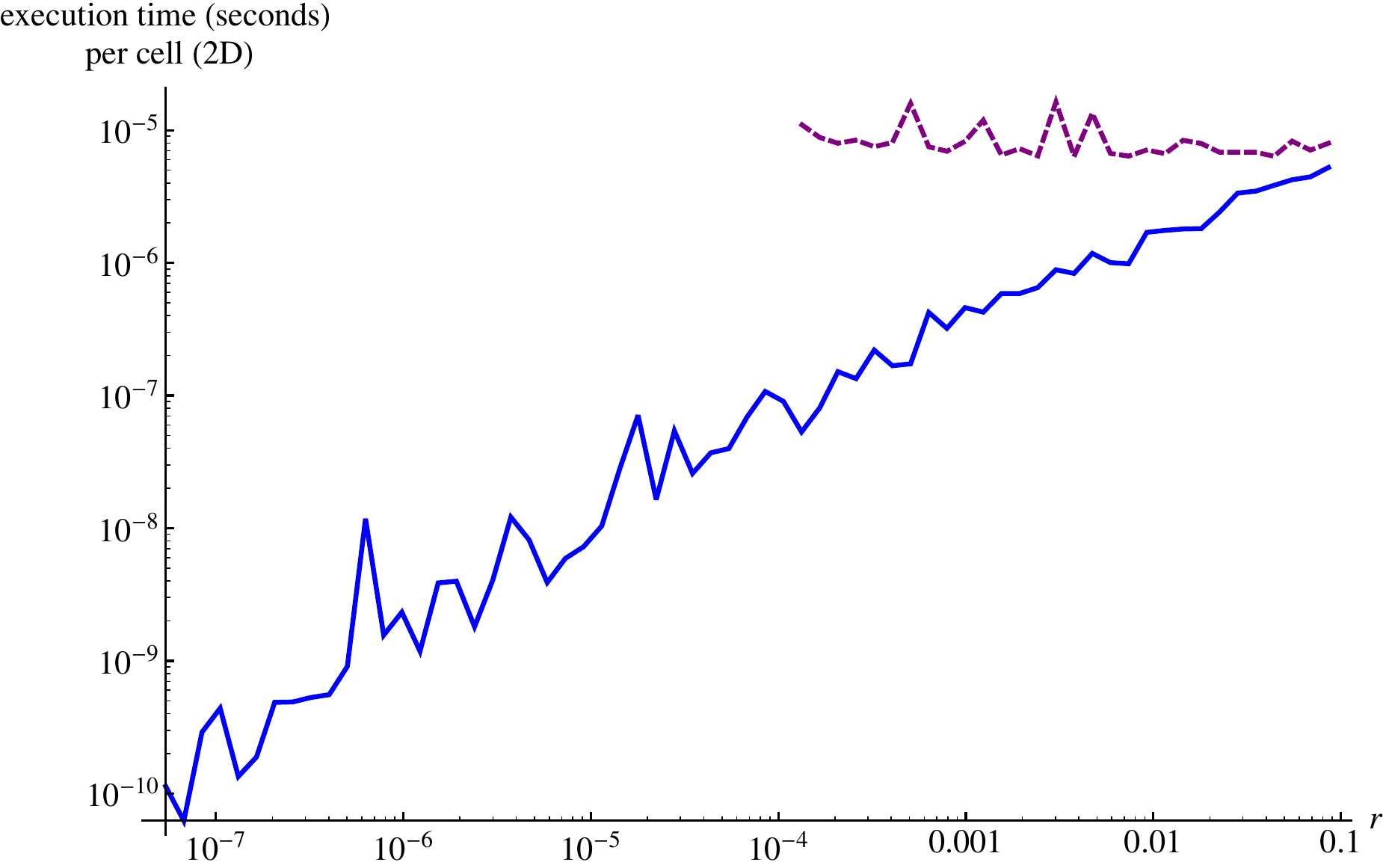}
\caption{Mean execution time per cell to find the first collision in a 2D square Lorentz  gas, for the classical (dashed curve) and efficient (solid curve) algorithms, as a function of disc radius, 
$r$.}
\label{fig:exectimepercell2D}
\end{figure}

\begin{figure}
\centering
\includegraphics [width=260pt]{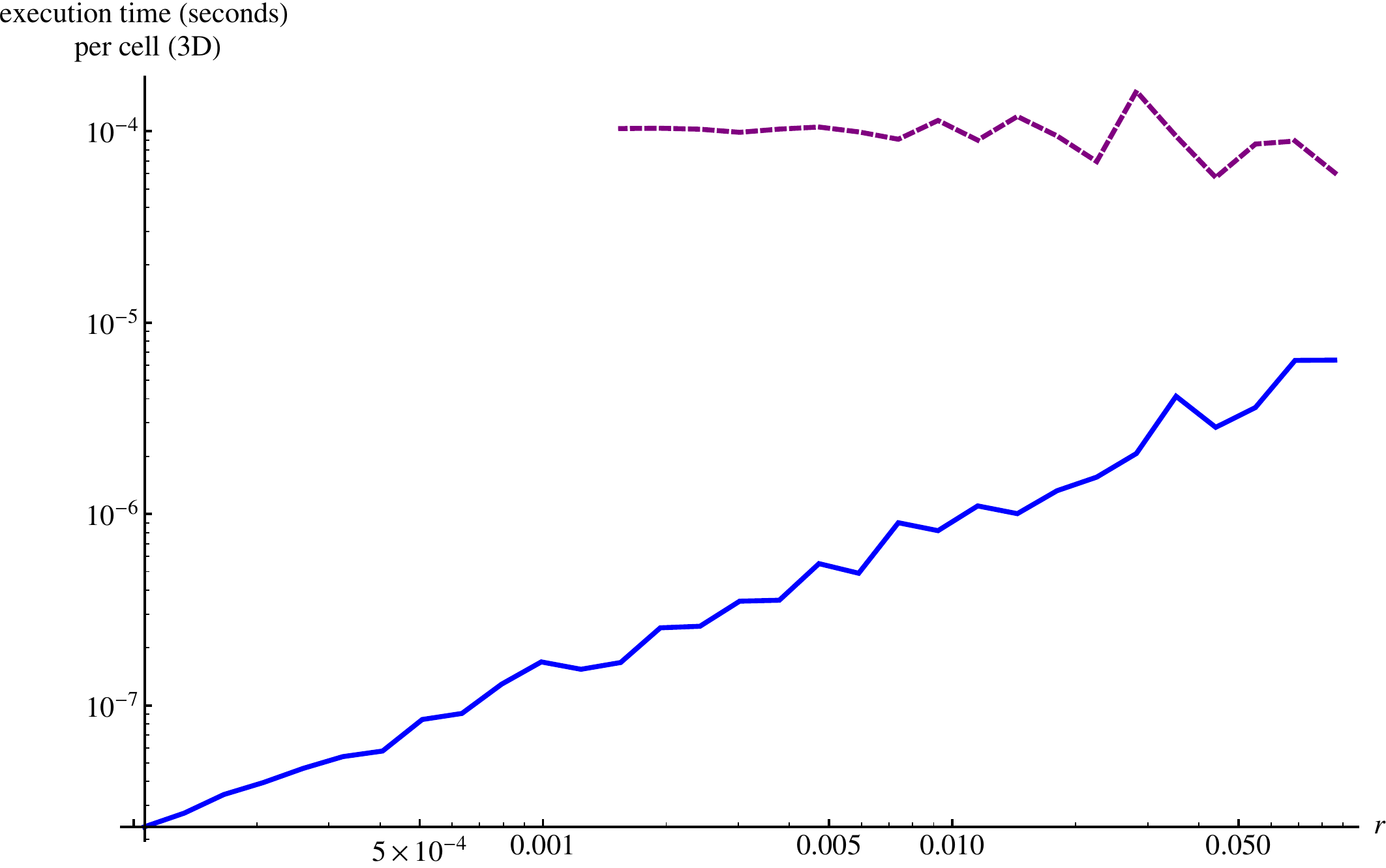}
\caption{Mean execution time per cell to find the first collision in the 3D cubic Lorentz gas, for the classical (dashed curve) and efficient (solid curve) algorithms, as a function of sphere radius, $r$.}
\label{fig:exectimepercell3D}
\end{figure}

Similarly, we calculated the execution time per cell as a function of the obstacle radius, for both the 2D and 3D efficient algorithms, with comparison to the corresponding classical algorithms; see Figures~\ref{fig:exectimepercell2D} and \ref{fig:exectimepercell3D}. Since the classical algorithms use periodic boundary conditions, the time per cell is basically constant, independent of the obstacle radius. 
For small radii, we again observe the efficiency of the new algorithms.

\subsection{Asymptotic complexity of the classical and efficient algorithms}
The scaling of the complexity of the classical algorithm as $1/r$ may be explained as follows.
The distance that a particle travels before it collides with an obstacle, i.e.~the free path length, is a function of obstacle size: the smaller the obstacles, the longer the free paths. 

In periodic Lorentz gases, there is a simple formula for the mean free path between consecutive collisions, $\tau$,  that arises from geometrical considerations \cite{ChernovFreePathJSP1997}: it is, up to a dimension-dependent constant, the ratio of the volume of the available space outside the obstacles to the surface area of the obstacles. For the square 2D Lorentz gas with discs of radius $r$, we have
\begin{equation}
\tau(r) = \frac{1 - \pi r^{2}}{2r},
\end{equation}
with asymptotics $r^{-1}$ for small $r$. Since the classical algorithm must cross this distance at speed $1$, it takes time proportional to $1/r$, as we find numerically. In applying this algorithm, approximately $1/r$ quadratic equations and four times as many linear equations must be solved.

For the simple cubic Lorentz gas in 3D with spheres of radius $r$, we have
\begin{equation}
\tau(r) = \frac{1 - \frac{4}{3} \pi r^{3}}{\pi r^2},
\end{equation}
with asymptotics $r^{-2}$, which is not far from our numerical results.

On the other hand, the efficient algorithm checks only one quadratic equation, and around 
$2r^{-1/2}$ linear equations, giving an upper bound of $r^{-1/2}$ for the complexity. (This calculation uses Hurwitz's theorem.) Numerically, it turns out to be significantly more efficient than that.

\subsection{Free flight distribution}
As an example application of our algorithm, we measure the distribution of free flight lengths for the first collision for certain systems studied by Marklof and Str\"ombergsson \cite{marklof2014power}. They studied $N$ incommensurable, overlapping periodic Lorentz gases in the Boltzmann--Grad limit, $r \to 0$, and proved that the asymptotic decay of the probability density for free flights in that system is $\sim \ell^{-N-2}$. It follows that the asymptotic density of the \emph{first} free flight should be 
$\rho(\ell) \sim \ell^{-N-1}$.

\begin{figure}
\centering
\includegraphics*[width=280pt]{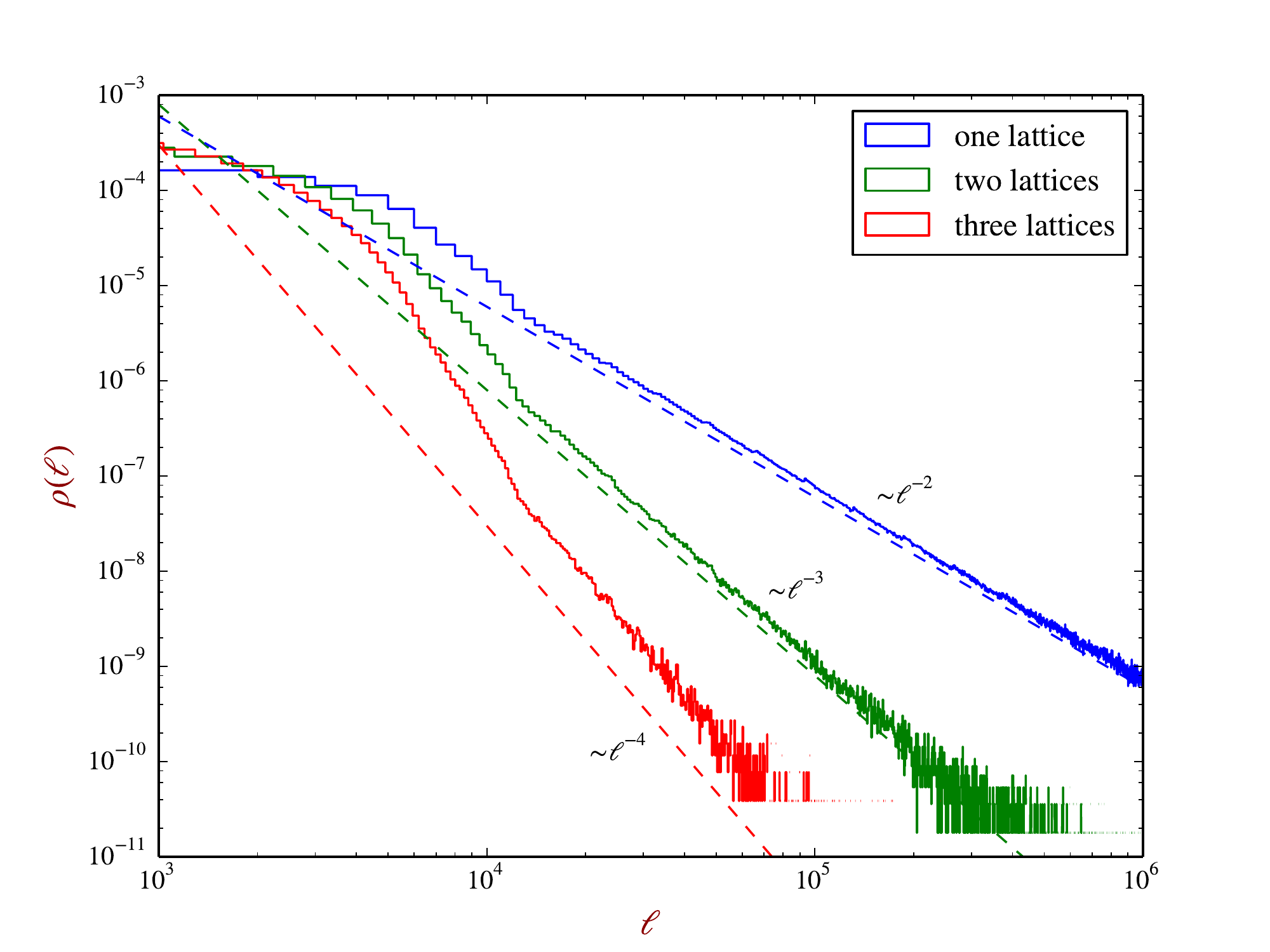}%{Free-Fligth-Rot.pdf}
\caption{Probability density of the first free flight for two and three incommensurable, overlapping periodic Lorentz gases with angles $\pi/5$ and $\pi/7$; a total of $10^{8}$ initial conditions was used. The results for a single lattice are shown for comparison. The dashed lines and labels show the theoretical asymptotics.}
\label{fig:free-flights}
\end{figure}

Figure~\ref{fig:free-flights} shows our numerical results for this distribution in the case of two and three overlapping lattices, compared to the asymptotic decay given by the rigorous result of \cite{marklof2014power}. 
To obtain this plot, we fixed the radius as $r=10^{-4}$ and calculated free flights for a given initial condition for a 2D lattice, and for the same lattice rotated by $\pi/5$ and $\pi/7$, respectively. The first free flight for each lattice is calculated separately, and the minimum of those results is then taken to give the first free flight for the superposition of either two or three incommensurable lattices. 
The distributions obtained numerically do indeed follow the power laws predicted.  Naturally, it becomes increasingly difficult to obtain the asymptotic behaviour of the densities as the number of lattices increases.

\section{Extension to general periodic lattices}

So far, we have restricted attention to spherical obstacles on simple cubic lattices. In this section, we will show how to deal with \emph{arbitrary} periodic crystal lattices. Such lattices consist of a \emph{basis} (finite collection) of different spheres (atoms), in unit cells of a Bravais lattice; see, e.g., \cite{AshcroftMermin}. 

This may be considered as the superposition of distinct Bravais lattices, one for each of the distinct atoms in the basis. Thus the efficient algorithm may be used separately for each such lattice, and then we take the minimum time to determine the next collision. In this way, we can now restrict attention to simulating a  Bravais lattice with a single atom per unit cell. For simplicity we will describe the method in 2D; the 3D case is similar.

A Bravais lattice in 2D is the set of points given by linear combinations of the form $a_{1} \vec{u}_{1} + a_{2} \vec{u}_{2}$ of vectors $\vec{u}_{i}$  defining the directions of the lattice, where the $a_{i}$ are integers. We pass from the square lattice to the oblique lattice by applying the transformation matrix 
$\mathsf{M}_\mathrm{so}$, defined such that its columns are the vectors $\vec{u}_{i}$:
\begin{equation}
\mathsf{M}_\mathrm{so} := (\vec{u}_1 | \vec{u}_2).
\end{equation}

\begin{figure}
\centering
\includegraphics*[width=300pt]{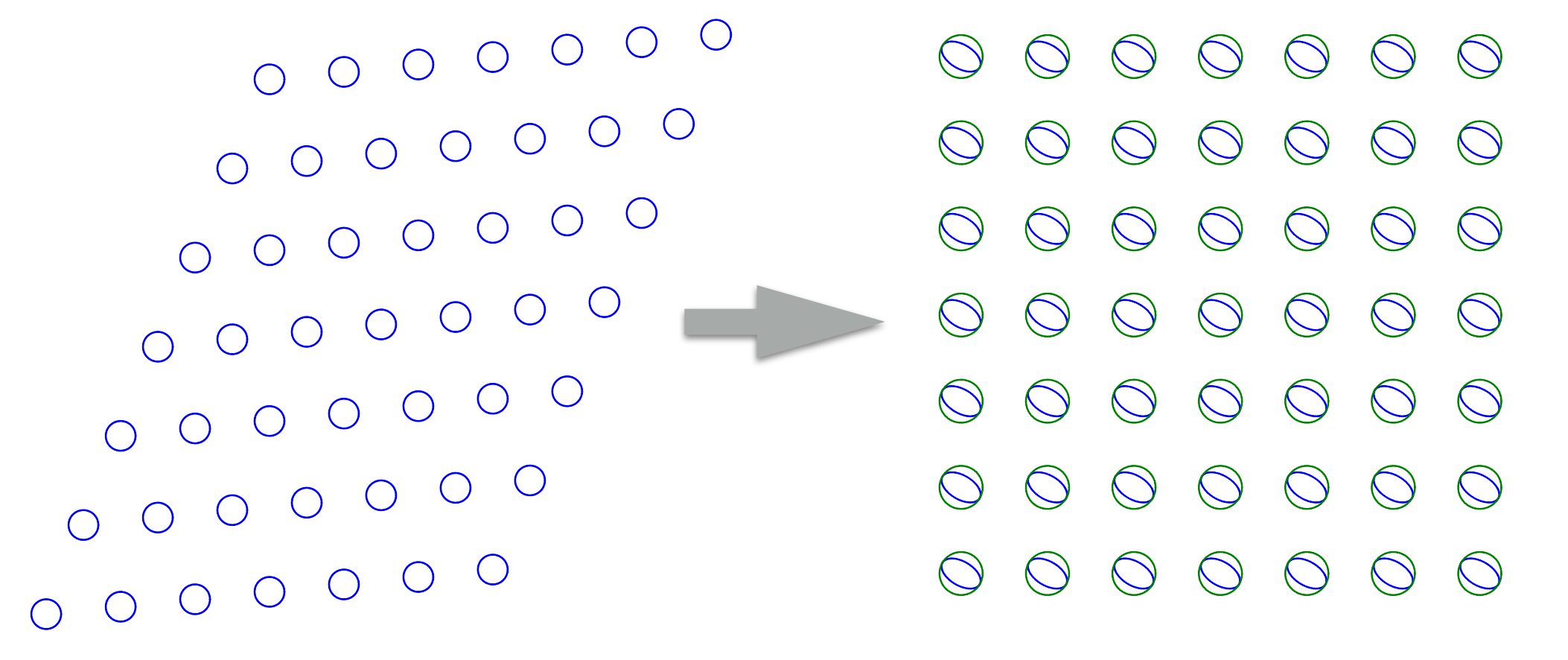}
\caption{Effect of applying the transformation $\mathsf{M}_\mathrm{os}$ to an oblique lattice of discs (left). The result is a square lattice of ellipses; circumscribed circles are also shown (right).
}
\label{fig:transformation}
\end{figure}

To transform back from the Bravais lattice to the square lattice, we apply the inverse transformation 
$\mathsf{M}_\mathrm{os} := \mathsf{M}_\mathrm{so}^{-1}$.
Starting from circular obstacles of radius $r$ in the Bravais lattice and applying $\mathsf{M}_\mathrm{os}$ gives one obstacle per unit cell at integer coordinates in the square lattice. However, this stretches the shape of the resulting obstacles into ellipses, as follows from the singular-value decomposition (SVD) of $\mathsf{M}_\mathrm{os}$; see, e.g., \cite{TrefethenBau}. 
The semi-major axis of the resulting ellipses is $r' = r \sigma_1$, where $\sigma_1$ is the first singular value of $\mathsf{M}_\mathrm{os}$. 
We circumscribe the resulting ellipse by a circular obstacle of radius $r'$, giving a standard square periodic Lorentz gas, suitable for analysis using the corresponding efficient 2D algorithm; see Figure~\ref{fig:transformation}.

Starting from a given initial condition $\vec{x}_{0}$, $\vec{v}_0$ in the Bravais lattice we wish to simulate, we transform these to 
$\vec{x}'_{0} := \mathsf{M}_\mathrm{os} \cdot \vec{x}_{0}$ and 
$\vec{v}'_{0} := \mathsf{M}_\mathrm{os} \cdot \vec{v}_{0}$
in the square lattice. 
We then apply the efficient algorithm in the square lattice to obtain a proposed disc or sphere with integer coordinates $\vec{n}$. These coordinates are mapped to the oblique lattice, giving a proposed disc or sphere with coordinates $\vec{n}' := \mathsf{M}_\mathrm{so} \cdot \vec{n}$.  We must check, however, if this is a true collision with the obstacle at $\vec{n}'$ using the classical algorithm, since the proposed collision with a disc in the square lattice may not actually hit the true elliptical obstacle there. If it is not a true collision, then we move to the next cell and continue; if it is a true collision, we calculate the new post-collision velocity. 

Provided the transformation $\mathsf{M}_\mathrm{so}$ does not stretch the obstacles too much, and the radius is small, this algorithm will still be very efficient.

%
%As an example, consider the 2D oblique Bravais lattice defined by vectors 
%$\vec{a_1} = {a_1 \choose 0}$ and $\vec{a_2} = {a_{21} \choose a_{22}}$, and obstacles of radius $r$, where the particle has initial position $\vec{x_0}$ and initial velocity $\vec{v_0}$. The transformation from the square lattice to the oblique lattice is
%$\mathsf{M}_\mathrm{so} = 
%\left( \begin{array}{cc}
%	a_1 & a_{21}\\
%	0 & a_{22}
%\end{array} \right)$
%and from oblique to square is
%$\mathsf{M}_\mathrm{os} = \mathsf{M}_\mathrm{so}^{-1} = 
%\left( \begin{array}{cc}
%\frac{1}{a_1} & 0\\
%-\frac{a_{21}}{a_1 a_{22}} & \frac{1}{a_{22}}
%\end{array} \right)$.
%In the square lattice, the transformed obstacle in each cell is an ellipse. The minimal radius $r'$ of a circle that circumscribes the ellipse is the semi-major axis, given by $r' = r \sigma_1$, where $\sigma_1$ is the first singular value of $\mathsf{M}_\mathrm{os}$. in this case is
%\begin{equation}
%r' = \frac{r}{\sqrt{2}} \sqrt{a_1^2 + a_{21}^2 + a_{22}^2 + \sqrt{(a_1^2 - a_{21}^2 - a_{22}^2)^2 + 4 a_1^2 a_{21}^2}}.
%\end{equation}
%We apply our efficient algorithm to the square lattice with obstacles of radius $r'$ and initial position and velocity $\mathsf{M}_\mathrm{os} \vec{x_0}$ and $\mathsf{M}_\mathrm{os} \vec{v_0}$. If a collision is found, we check if it is real by the classical method in the original (oblique) lattice by using the initial conditions $\vec{x_0}$, $\vec{v_0}$ and the obstacle corresponding (by the transformation $\mathsf{M}_\mathrm{so}$) to the one found in the square lattice.

Finally, non-spherical obstacles may be dealt with in a similar way, using a circumscribed circular or spherical obstacle. In this way, we may simulate completely general crystal lattice structures.

\section{Conclusions}  

We have introduced efficient algorithms to simulate periodic Lorentz gases in two and three dimensions, that work particularly well when the obstacles are small. We have compared the efficiency of these algorithms with the standard ones, showing that the relative efficiency indeed increases very fast in 2D and fast in 3D, and we have shown a sample application to calculate free flight distributions near the Boltzmann--Grad limit.

We have also shown how to extend our methods to arbitrary crystal lattices. 
The extension of the 3D algorithm to higher dimensions and applications are  in progress. 

\section{Acknowledgements}  
We thank Michael Schmiedeberg for useful comments and discussions about the algorithms, and the anonymous referees for their insightful remarks. 
ASK received support from the DFG within the Emmy Noether program (grant Schm 2657/2). 
NK is the recipient of a DGAPA-UNAM postdoctoral fellowship. 
DPS acknowledges financial support from CONACYT grant CB-101246 and DGAPA-UNAM PAPIIT grants IN116212 and IN117214. 

\vspace*{10pt}

\bibliographystyle{unsrt}
\bibliography{efficient_Lorentz}

\end{document}